# Molecular Dynamics Studies of Dislocations in CdTe Crystals from a New Bond Order Potential

X. W. Zhou,[a,*] D. K. Ward,[b] B. M. Wong,[c] F. P. Doty,[b] and J. A. Zimmerman[a]

**E-mail:** xzhou@sandia.gov

[a]Mechanics of Materials Department, Sandia National Laboratories, Livermore, California 94551, USA

[b]Radiation and Nuclear Detection Materials and Analysis Department, Sandia National Laboratories, Livermore, California 94551, USA

[c]Materials Chemistry Department, Sandia National Laboratories, Livermore, California 94551, USA

*Abstract*—**$Cd_{1-x}Zn_xTe$ (CZT) crystals are the leading semiconductors for radiation detection, but their application is limited by the high cost of detector-grade materials. High crystal costs primarily result from property non-uniformity that causes low manufacturing yield. While tremendous efforts have been made in the past to reduce Te inclusions / precipitates in CZT, this has not resulted in an anticipated improvement in material property uniformity. Moreover, it is recognized that in addition to Te particles, dislocation cells can also cause electric field perturbation and the associated property non-uniformity. Further improvement of the material, therefore, requires that dislocations in CZT crystals be understood and controlled. Here we use a recently developed CZT bond order potential to perform representative molecular dynamics simulations to study configurations, energies, and mobilities of 29 different types of possible dislocations in CdTe (i.e., x = 1) crystals. An efficient method to derive activation free energies and activation volumes of thermally activated dislocation motion will be explored. Our focus gives insight into understanding important dislocations in the material, and gives guidance toward experimental efforts for improving dislocation network structures in CZT crystals.**

I. INTRODUCTION

While CdTe-based CZT ($Cd_{1-x}Zn_xTe$) crystals are the leading semiconductor compounds for radiation-detection applications [1-4], the wide-spread deployment of CZT detectors has been limited by the high cost (due to a low manufacturing yield) of detector-grade materials. Property non-uniformity has been the major cause for both poor performance and a low yield of usable portions of ingots [1]. It has long been established that micron-scale defects such as tellurium inclusions / precipitates affect carrier transport and uniformity [1,5,6]. However, extensive previous efforts to reduce Te particles have not resulted in an anticipated improvement in CZT property uniformity. One special structural feature of the CZT crystals is that they are extremely soft and, therefore, always develop a high density of dislocations during synthesis [7-9]. Experiments have shown that when these dislocations are organized into a network of sub-grains, they can directly affect charge carriers [10,11] to cause electric field perturbations. The sub-grain boundaries can also serve as nucleation sites for tellurium precipitates [1]. Clearly, eliminating or at least controlling dislocation sub-grain structures holds great potential to further improve CZT crystals.

Many types of dislocations may be present in CZT crystals, including shuffle dislocations, glide dislocations, α and β dislocations, partial and perfect dislocations, edge, screw, and mixed dislocations, and misfit dislocations. The understanding of the behavior of the various combinations of these dislocations in conjunction with developing methods for improving the corresponding dislocation network structures are therefore challenging. The use of predictive, high-fidelity atomistic simulations can help screen out the structure-limiting dislocation types. Here we use a recently developed CZT bond order potential (BOP) [12-14][*] in LAMMPS (Large-scale Atomic/Molecular Massively Parallel Simulator) [15,16] to perform selected molecular dynamic (MD) simulations to determine configurations, energies, and mobility of 29 different types of dislocations in CdTe crystals. An efficient method to derive activation free energy and activation volume of thermally activated dislocation motion will be explored. Our focus is to identify and validate important dislocations and dislocation phenomena in the material, which can reduce future experimental efforts to improve the dislocation network structures.

## II. INTERATOMIC POTENTIAL

The key to high-fidelity MD simulations of dislocations is an interatomic potential that is transferrable to the dislocated regions where the local atomic environment is significantly different from that of the perfect bulk crystal. Details of the CZT bond order potential applied here are omitted as they are very complex and are fully described previously [12-14]. It is demonstrated that the CZT BOP captures well the property trends of a variety of elemental and compound structures / phases (with local environment varying from coordination 2 to 12) [12-14], simulates correctly the crystalline growth of stoichiometric compounds during both vapor deposition (under non-stoichiometric vapor fluxes) [17] and melt-growth [18] processes, and predicts precisely the configuration and density of misfit dislocations in vapor deposited multilayers [19]. Note that the correct simulation of growth, especially under a non-stoichiometric environment, is a very strong validation of the transferability of a potential as the process involves chemical reactions and samples a variety of configurations and chemistries statistically formed on a growth surface. For this reason, the CZT BOP is expected to produce reasonable results regarding dislocations.

## III. DEFINITION OF DISLOCATION TYPES

Different types of dislocations can be constructed on various planes in a zinc-blende crystal. For instance, mobile dislocations usually lie on the {111} slip planes, and misfit dislocations in multilayered films are often seen to lie on interfacial planes [19], which are not necessarily parallel to {111}. Here we only study dislocations that are on the {111} slip planes, with Burgers vector $\vec{b}$ being either partial <112>a/6 or perfect <110>a/2.

---

[*]The CZT ternary BOP [14] is an extension of the CdTe binary BOP [12] so they are equivalent if used only for CdTe.

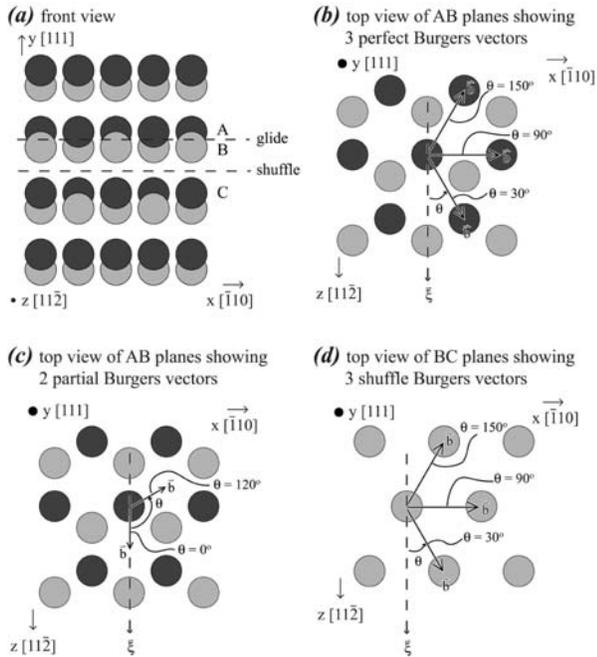
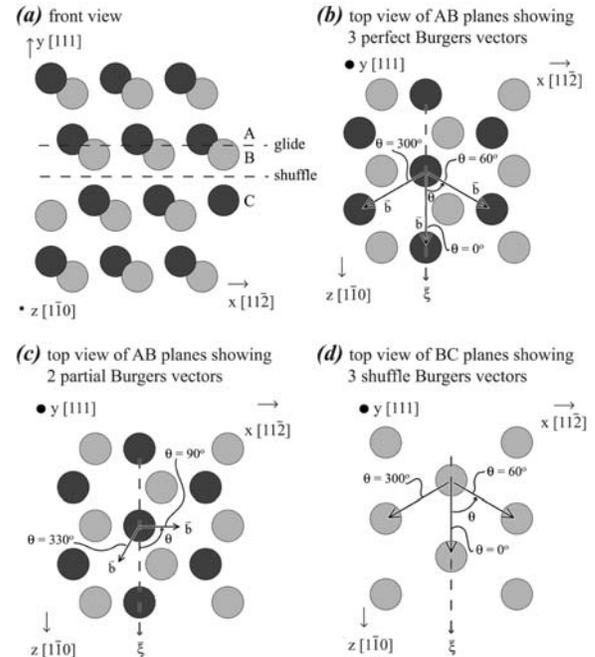

Fig. 1. Geometry of the edge coordinate system. (a) front projection of $(11\bar{2})$ planes; (b) top projection of two (111) planes A and B showing glide perfect dislocations; (c) top projection of two (111) planes A and B showing glide partial dislocations; and (d) top projection of two (111) planes B and C showing shuffle dislocations.

Fig. 2. Geometry of the screw coordinate system. (a) front projection of $(1\bar{1}0)$ planes; (b) top projection of two (111) planes A and B showing glide perfect dislocations; (c) top projection of two (111) planes A and B showing glide partial dislocations; and (d) top projection of two (111) planes B and C showing shuffle dislocations.

Assuming that the dislocation line direction $\vec{\xi}$ aligns with the z direction and the slip occurs on the x-z plane, our definition of different types of dislocations is shown in Figs. 1 and 2, where dark and light circles distinguish metal (Cd or Zn) and Te atoms, respectively. Dislocations defined in Fig. 1 are based upon an edge dislocation coordinate system shown in Fig. 1(a) where the dislocation line $\vec{\xi}$ is perpendicular to one of the Burgers vectors <110>a/2 (x axis). Dislocations defined in Fig. 2 are based upon a screw dislocation coordinate system shown in Fig. 2(a) where $\vec{\xi}$ is parallel to one of the perfect Burgers vectors <110>a/2 (z axis). Figs. 1(a) and 2(a) indicate that dislocation slip can occur at two different locations, one between two closely separated planes A and B (where the inter-plane bonds are not perpendicular to the planes), and the other between the more widely separated planes B and C (where the inter-plane bonds are perpendicular to the planes). According to convention, dislocations between A and B are called glide dislocations, and dislocations between planes B and C are called shuffle dislocations.

Burgers vectors of glide dislocations can be examined from a top projection of two isolated planes A and B. Figs. 1(b) and 2(b) use such a projection to explore perfect dislocation Burgers vectors (<110>a/2). It can be seen from Fig. 1(b) that in the edge coordinate system, there are three non-equivalent perfect Burgers vectors that can be represented by the angle θ between $\vec{b}$ and $\vec{\xi}$, i.e., θ = 30°, 90°, 150°. There are other perfect Burgers vectors, but they are equivalent to the ones listed due to crystal symmetry. For instance, the θ = 330° dislocation is equivalent to the θ = 30° dislocation. Similarly, it can be seen from Fig. 2(b) that in the screw coordinate system, there are three more non-equivalent perfect Burgers vectors θ = 0°, 60°, 300°.

Figs. 1(c) and 2(c) use the same top projection of two isolated planes A and B to explore partial dislocation Burgers vectors (<112>a/6). It can be seen from Fig. 1(c) that in the edge coordinate system, there are two non-equivalent partial Burgers vectors θ = 0°, 120°. Similarly, it can be seen from Fig. 2(c) that in the screw coordinate system, there are two more non-equivalent partial Burgers vectors θ = 90°, 330°.

Burgers vectors of shuffle dislocations can be examined from top projection of two isolated planes B and C. Figs. 1(d) and 2(d) use such a projection to explore shuffle dislocation Burgers vectors. Here the light and dark atoms in the B and C planes are aligned in the projection (i.e., only light atoms are seen), and hence, there are no partial dislocation Burgers vectors for shuffle dislocations. As indicated in Fig. 1(d) for the edge coordinate system, we explore three shuffle dislocation Burgers vectors θ = 30°, 90°, 150°, which are the same as those in the perfect glide dislocation case shown in Fig. 1(b). As indicated in Fig. 2(d) for the screw coordinate system, we explore three more shuffle dislocation Burgers vectors θ = 0°, 60°, 300°, which are the same as those in the perfect glide dislocation case shown in Fig. 2(b).

Dislocations shown in Figs. 1 and 2 always occur between a metal (Cd or Zn) and a Te {111} planes. If a dislocation has an edge component, then the dislocation core can have extra metal atoms or extra Te atoms. According to convention, a dislocation with more metal atoms at its core is called an α dislocation, and a dislocation with more Te atoms at its core is called a β dislocation.

According to the discussion above, we will consider a combination of glide and shuffle dislocations, partial and perfect dislocations, various Burgers vectors, and α and β dislocations. These constitute 29 different types of dislocations as listed in Table I.

## IV. STACKING FAULT ENERGY

A CdTe crystal system with 336 ($11\bar{2}$) planes in the x direction, 60 (111) planes in the y direction, and 28 ($1\bar{1}0$) planes in the z direction is used to calculate the (111) stacking fault energy $\gamma_{sf}$. The stacking fault is created by shifting the upper part of a perfect crystal a partial dislocation vector with respect to the lower part (at a glide dislocation slip plane between planes A and B, see Fig. 1 or 2). Molecular statics (MS) energy minimization simulations are used to calculate relaxed energies of both the perfect and the stacking faulted crystals under the periodic boundary conditions in both x and z directions and a free boundary condition in y direction. The stacking fault energy is then calculated as $\gamma_{sf} = (E_{sf} - E_0)/(L_x \cdot L_z)$, where $E_{sf}$ and $E_0$ are respectively the relaxed energies for stacking faulted and perfect crystals, and $L_x$ and $L_z$ are respectively the system dimensions in x and z directions. Note that the surface energy in the y direction does not affect the calculation as it exists for both stacking faulted and perfect crystals and therefore cancels out. We find that $\gamma_{sf} \approx 0$ in our calculations. In the literature, CdTe is known to have a low stacking fault energy around 10 mJ/m$^2$ [20,21].

## V. DISLOCATION LINE ENERGIES

Linear elastic continuum theory cannot describe dislocation core structures. Atomistic simulations allow direct studies of dislocation core structures. This in turn allows dislocation energies, i.e., the energy difference between a dislocated crystal and a perfect CdTe crystal containing the same number of atoms, to be calculated as a function of dislocation core radius. The key for the calculations is to use the initial dislocation configurations that can be fully relaxed during MD and MS energy minimization simulations. Different methods have been tried to create initial dislocations. We find that an initial dislocation created by simply positioning atoms according to the displacement field of the linear elastic continuum solution does not always fully relax at the core. Instead, a dislocation created by shearing the relevant portions of the crystals using MD simulations is found to normally relax to lower energies than the other methods. In particular, as will be described below, an MD scheme illustrated in Fig. 3 is effective in creating partial, un-dissociated perfect, and shuffle dislocations, and an MD scheme illustrated in Fig. 4 is effective in creating dissociated perfect dislocations. These schemes enable atom positions to be strictly determined from atom interactions and also best reflects realistic scenario that dislocations are created under local shear stresses.

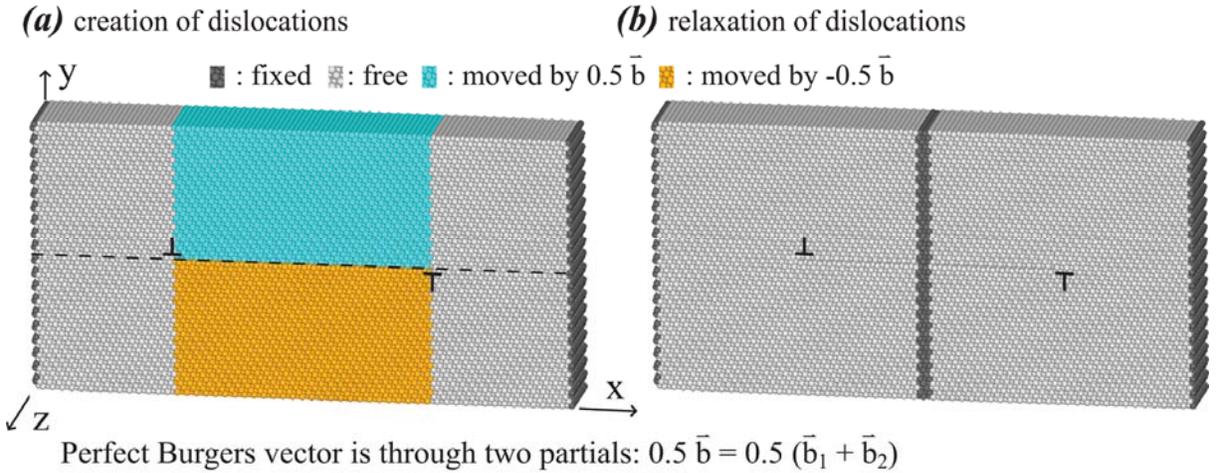

Fig. 3. Line energy model for partial, un-dissociated perfect, and shuffle dislocations. (a) MD simulations to create dislocations; and (b) MD and MS simulations to relax dislocations.

The computational system used is periodic in the x and z axes with free surfaces along the y axis. For the un-dissociated dislocations (including partial and shuffle dislocations) shown in Fig. 3, the system is divided into different regions as indicated by blue (middle upper), orange (middle lower), black (left and right), and white (the rest) colors. With the black regions held fixed and the white regions free to relax, an MD simulation is first performed to uniformly move the blue region by half of the Burgers vector $0.5\vec{b}$ and the orange region by negative half of the Burgers vector $-0.5\vec{b}$. For partial dislocations, the movement of the blue and orange regions is completed directly over a total simulated time of 0.1 ns. For perfect dislocations, the movement is through two stages over a total simulated time of 0.2 ns to follow the lowest energy path of two partials: i.e., $0.5(\vec{b}_1+\vec{b}_2)$ ($=0.5\vec{b}$) for the blue region and $-0.5(\vec{b}_1+\vec{b}_2)$ ($=-0.5\vec{b}$) for the orange region. In addition, the system temperature is uniformly decreased from 300 K to 10 K during simulation. This MD procedure creates two dislocations with opposite Burgers vectors (also note that if the dislocations have an edge component, then one is an α and the other one is a β dislocation) as shown in Fig. 3(a).

For the dissociated perfect dislocations shown in Fig. 4, the system is divided into different regions as indicated by light and dark blue (middle upper), light and dark orange (middle lower), black (left and right), and white (the rest) colors. With the black regions held fixed and the white regions free to relax, an MD simulation is first performed to uniformly move the light and dark blue region by half of a Shockley partial Burgers vector $0.5\vec{b}_1$ and the light and dark orange region by negative half of the same Shockley partial Burgers vector $-0.5\vec{b}_1$ over a total simulated time of 0.1 ns. A second MD simulation is then performed to move the dark blue region by half of the other Shockley partial Burgers vector $0.5\vec{b}_2$ and the dark orange region by negative half of the same Shockley partial Burgers vector $-0.5\vec{b}_2$ over another total simulated time of 0.1 ns. In addition, the system temperature is uniformly decreased from 300 K to 10 K during these two simulations. This MD procedure creates the same two perfect dislocation Burgers vectors as shown in Fig. 3(a) except that each perfect dislocation is dissociated into two Shockley partials

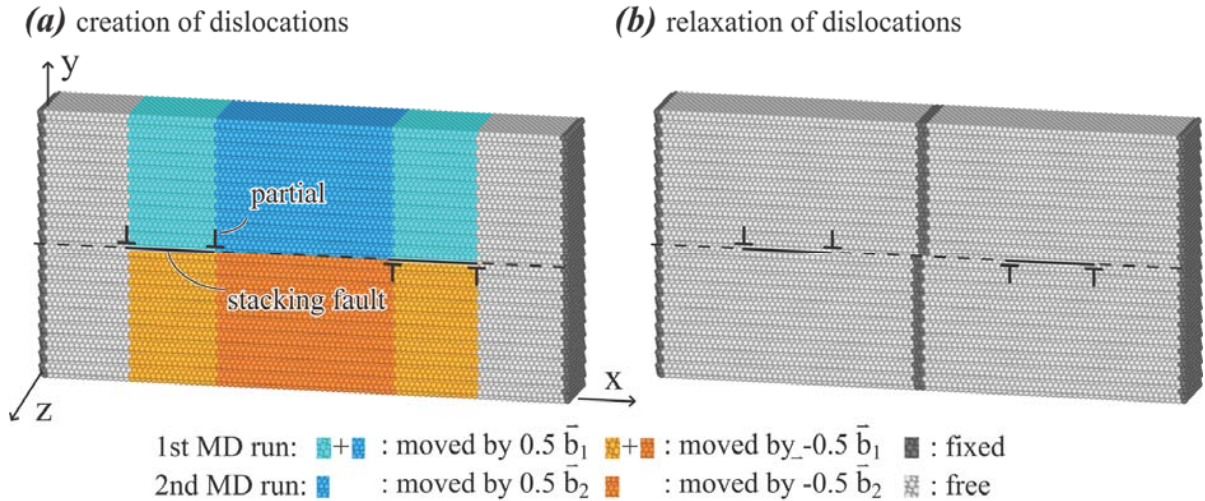

Fig. 4. Line energy model for dissociated perfect dislocations. (a) MD simulations to create dislocations; and (b) MD and MS simulations to relax dislocations.

bounding a stacking fault band as shown in Fig. 4(a).

Once dislocations are created in the MD simulations, the blue and orange regions are released. A narrow region in the middle of the system is allocated as shown by the black color in Figs. 3(b) and 4(b). For perfect dislocations, this narrow region remains a perfect crystal because the blue region is moved by a perfect Burgers vector with respect to the orange region. For partial dislocations, the narrow middle region contains a stacking fault. With left, right, and middle black regions fixed, an MS simulation is then conducted to relax the dislocations.

In this work, we use an edge dislocation geometry containing 97 ($\bar{1}10$) planes in the x direction, 60 (111) planes in the y direction, and 24 ($11\bar{2}$) planes in the z direction and a screw dislocation geometry containing 168 ($11\bar{2}$) planes in the x direction, 60 (111) planes in the y direction, and 14 ($1\bar{1}0$) planes in the z direction to create various perfect and dislocated computational crystals. Relaxed energies of both dislocated and perfect crystals are calculated. To separate α and β dislocation effects, the total energy is split to the left and right parts respectively by summing up the atomic energies of the left and right halves of the systems. The line energies of the left and the right dislocation, $\Gamma_i$, is then calculated as $\Gamma_i = (E_{d,i} - E_{0,i} - w_i \cdot L_z \cdot \gamma_{sf}) / L_z$, where $E_d$ and $E_0$ are the relaxed energies of the dislocated and the perfect crystals respectively, $L_z$ is dislocation length (= system dimension in the z axis), $\gamma_{sf}$ is stacking fault energy, w is stacking fault band width, and the subscript i (= l, r) indicates the left and the right dislocations. The results obtained are summarized in Table. I.

TABLE I

DISLOCATION LINE ENERGIES IN CdTe CRYSTALS.

| $\bar{\xi}$ | $\bar{b}$ | θ (°) | type | | energy (eV/Å) |
|---|---|---|---|---|---|
| \multicolumn{6}{c}{7 glide partial dislocations} | | | | | |
| [$11\bar{2}$] | [$11\bar{2}$]/6 | 0 | α (=β) | | 0.38 |
| [$11\bar{2}$] | [$\bar{2}11$]/6 | 120 | α | | 0.63 |
| [$11\bar{2}$] | [$\bar{2}11$]/6 | 120 | β | | 0.63 |
| [$1\bar{1}0$] | [$11\bar{2}$]/6 | 90 | α | | 0.70 |
| [$1\bar{1}0$] | [$11\bar{2}$]/6 | 90 | β | | 0.70 |
| [$1\bar{1}0$] | [$1\bar{2}1$]/6 | 330 | α | | 0.44 |
| [$1\bar{1}0$] | [$1\bar{2}1$]/6 | 330 | β | | 0.45 |

| $\bar{\xi}$ | $\bar{b}$ | θ (°) | Shockley dissociations | | type | energy (eV/Å) |
|---|---|---|---|---|---|---|
| \multicolumn{7}{c}{11 glide perfect dislocations (dissociated)} | | | | | | |
| [$1\bar{1}0$] | [$1\bar{1}0$]/2 | 0 | [$1\bar{2}1$]/6 | [$2\bar{1}\bar{1}$]/6 | α (=β) | 0.69 |
| [$1\bar{1}0$] | [$10\bar{1}$]/2 | 60 | [$11\bar{2}$]/6 | [$2\bar{1}\bar{1}$]/6 | α | 1.38 |

| $\vec{\xi}$ | $\vec{b}$ | θ (°) | | | type | energy (eV/Å) |
|---|---|---|---|---|---|---|
| [1$\bar{1}$0] | [10$\bar{1}$]/2 | 60 | [11$\bar{2}$]/6 | [2$\bar{1}\bar{1}$]/6 | β | 1.38 |
| [1$\bar{1}$0] | [10$\bar{1}$]/2 | 300 | [1$\bar{2}$1]/6 | [$\bar{1}\bar{1}$2]/6 | α | 1.38 |
| [1$\bar{1}$0] | [0$\bar{1}$1]/2 | 300 | [1$\bar{2}$1]/6 | [$\bar{1}\bar{1}$2]/6 | β | 1.38 |
| [11$\bar{2}$] | [01$\bar{1}$]/2 | 30 | [11$\bar{2}$]/6 | [$\bar{1}$2$\bar{1}$]/6 | α | 1.05 |
| [11$\bar{2}$] | [01$\bar{1}$]/2 | 30 | [11$\bar{2}$]/6 | [$\bar{1}$2$\bar{1}$]/6 | β | 1.03 |
| [11$\bar{2}$] | [$\bar{1}$10]/2 | 90 | [$\bar{2}$11]/6 | [$\bar{1}$2$\bar{1}$]/6 | α | 1.60 |
| [11$\bar{2}$] | [$\bar{1}$10]/2 | 90 | [$\bar{2}$11]/6 | [$\bar{1}$2$\bar{1}$]/6 | β | 1.55 |
| [11$\bar{2}$] | [$\bar{1}$01]/2 | 150 | [$\bar{2}$11]/6 | [$\bar{1}\bar{1}$2]/6 | α | 1.04 |
| [11$\bar{2}$] | [$\bar{1}$01]/2 | 150 | [$\bar{2}$11]/6 | [$\bar{1}\bar{1}$2]/6 | β | 1.03 |

| 11 glide perfect dislocations (un-dissociated) ||||
|---|---|---|---|
| $\vec{\xi}$ | $\vec{b}$ | θ (°) | type | energy (eV/Å) |
| [1$\bar{1}$0] | [1$\bar{1}$0]/2 | 0 | α (=β) | 0.79 |
| [1$\bar{1}$0] | [10$\bar{1}$]/2 | 60 | α | 1.51 |
| [1$\bar{1}$0] | [10$\bar{1}$]/2 | 60 | β | 1.66 |
| [1$\bar{1}$0] | [10$\bar{1}$]/2 | 300 | α | 1.56 |
| [1$\bar{1}$0] | [0$\bar{1}$1]/2 | 300 | β | 1.53 |
| [11$\bar{2}$] | [01$\bar{1}$]/2 | 30 | α | 1.32 |
| [11$\bar{2}$] | [01$\bar{1}$]/2 | 30 | β | 1.16 |
| [11$\bar{2}$] | [$\bar{1}$10]/2 | 90 | α | 1.73 |
| [11$\bar{2}$] | [$\bar{1}$10]/2 | 90 | β | 1.77 |
| [11$\bar{2}$] | [$\bar{1}$01]/2 | 150 | α | 1.14 |
| [11$\bar{2}$] | [$\bar{1}$01]/2 | 150 | β | 1.15 |

| 11 shuffle dislocations ||||
|---|---|---|---|
| $\vec{\xi}$ | $\vec{b}$ | θ (°) | type | energy (eV/Å) |
| [1$\bar{1}$0] | [1$\bar{1}$0]/2 | 0 | α (=β) | 0.81 |
| [1$\bar{1}$0] | [10$\bar{1}$]/2 | 60 | α | 1.46 |
| [1$\bar{1}$0] | [10$\bar{1}$]/2 | 60 | β | 1.46 |
| [1$\bar{1}$0] | [0$\bar{1}$1]/2 | 300 | α | 1.49 |
| [1$\bar{1}$0] | [0$\bar{1}$1]/2 | 300 | β | 1.49 |
| [11$\bar{2}$] | [01$\bar{1}$]/2 | 30 | α | 1.05 |
| [11$\bar{2}$] | [01$\bar{1}$]/2 | 30 | β | 1.05 |
| [11$\bar{2}$] | [$\bar{1}$10]/2 | 90 | α | 1.70 |
| [11$\bar{2}$] | [$\bar{1}$10]/2 | 90 | β | 1.70 |
| [11$\bar{2}$] | [$\bar{1}$01]/2 | 150 | α | 1.07 |
| [11$\bar{2}$] | [$\bar{1}$01]/2 | 150 | β | 1.07 |

The relaxed configurations obtained from the energy minimization simulations were visualized. We find that of all the simulations using the initial perfect dislocations, the maximum splitting distance is only 25 Å after energy minimization. This confirms that these dislocations indeed remain approximately un-dissociated. Table I indicates that the screw ($\theta = 0°$) dislocations have lower energies than other dislocations, being 0.38 eV/Å, 0.69 eV/Å, 0.79 eV/Å, and 0.81 eV/Å for partial, dissociated perfect, un-dissociated perfect, and shuffle dislocations, respectively. Interestingly, the acute angle of the second lowest energy (shuffle, glide partial, glide perfect) dislocations all equals 30° (e.g., $\theta = 30°, 150°, 330°$). In addition, the energies of dissociated perfect dislocations are about 80% - 92% of the energies of un-dissociated dislocations, suggesting that the dissociated dislocations are more stable. However, the observation that the perfect dislocations remain approximately un-dissociated during MD and energy minimization simulations suggests an energy barrier of dissociation. The configuration of perfect dislocations will be examined further below in MD simulations of dislocation motion.

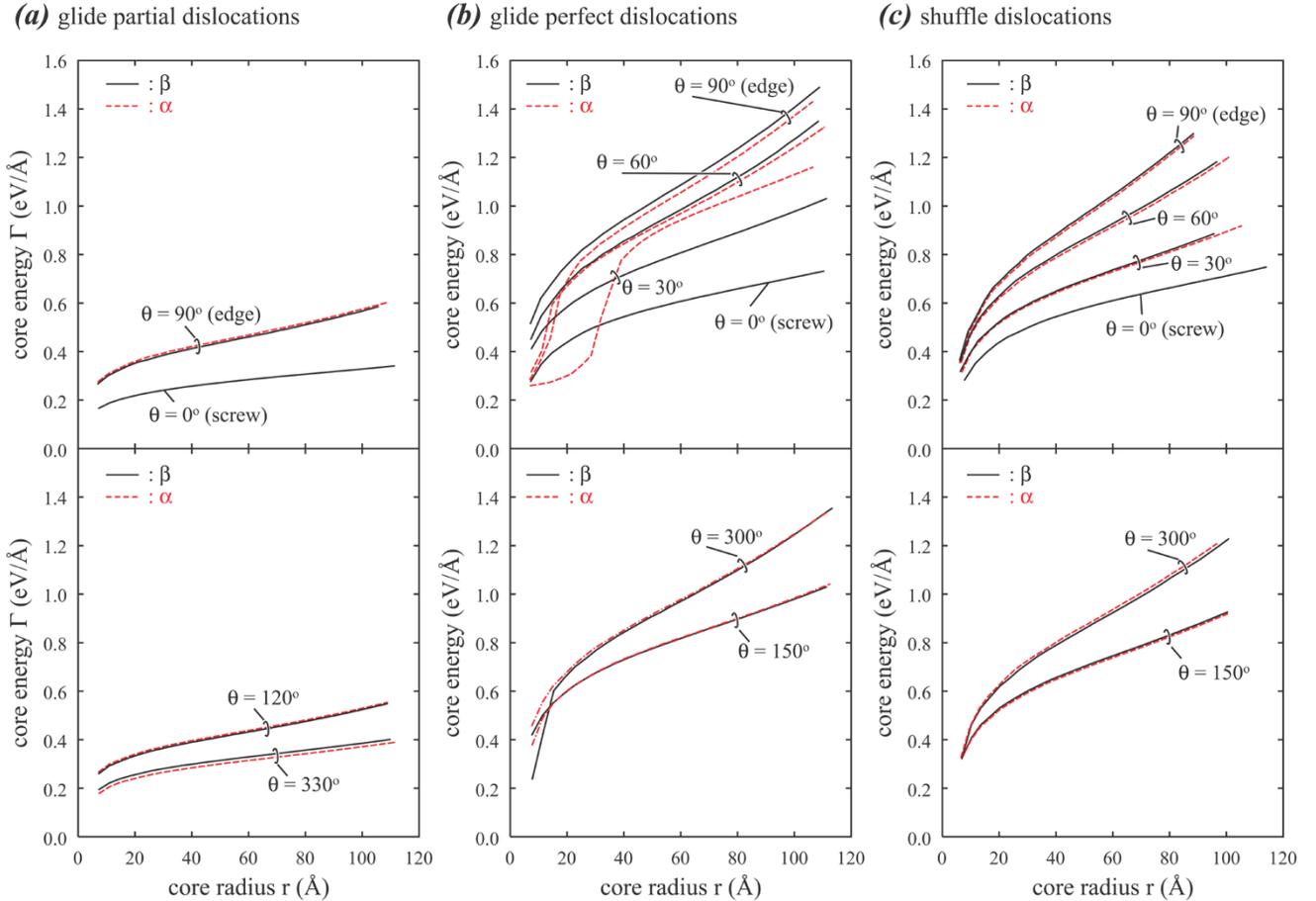

Fig. 5. Dislocation line energy as a function of core radius. (a) glide partial dislocations; (b) glide perfect dislocations; and (c) shuffle dislocations.

Dislocation line energy is not a constant material property, but rather increases with the material volume within a radius $r$

from the dislocation core. Dislocation line energy at a given radius r can be calculated using the same approach described above except that the energies of the perfect and dislocated crystals are taken as the sum of the atomic energies for all the atoms within the radius. The dislocation line energies as a function of r thus obtained are shown in Figs. 5(a)-5(c) for partial, (un-dissociated) perfect, and shuffle dislocations respectively.

Figs. 5(a)-5(c) indicates that dislocation line energies calculated with the atomistic models linearly increase with dislocation radius r when r is larger than 40 Å. This is consistent with the boundary conditions used in Figs. 3 and 4 where the black regions are held fixed during simulations. The dislocation core energies at small radii, say r < 30 Å, come from the relaxed dislocation core structures. In the present work, the system sizes have been chosen to be sufficiently large so that the predicted dislocation core structures are independent of the boundary conditions. As a result, our simulations can give accurate dislocation core energies at r < 30 Å that cannot be predicted from linear elastic theories. It can be seen from Figs. 5(a)-5(c) that dislocation energies sharply decrease when the radius is decreased. In addition, α and β dislocations have similar energy vs. r curves for partial and shuffle dislocations. While α and β dislocation energies are also similar for the perfect dislocations at large radii, they become different at the core. In particular, the α perfect dislocation core energies are lower than the β perfect dislocation core energies.

## VI. Dislocation Loop Evolution under Shear

One effective approach to simultaneously study the motion of multiple dislocations is to examine the evolution of a dislocation loop under a shear stress. The computational system consists of a CdTe crystal with 120 $(11\bar{2})$ planes in the x direction, 18 (111) planes in the y direction, and 68 $(1\bar{1}0)$ planes in the z direction. Periodic boundary conditions are used in x and z directions and a free boundary condition is used in y direction during the simulation. To create a dislocation loop, the system is divided into different regions as shown in Fig. 6(a): the black (top and bottom) regions, the blue (middle upper) hexagonal region, the orange (middle lower) hexagonal region, and the white (remaining) region. As plastic deformation usually proceeds through glide of dislocations, we focus on glide dislocations. Hence, we assume that the boundary between the upper and lower hexagons is between planes A and B as defined in Figs. 1(a) and 2(a). To create a realistic dislocation loop with a Burgers vector of $\vec{b}$, an MD simulation is performed to uniformly move the top black and blue regions by $0.5\vec{b}$ and the lower black and orange regions by $-0.5\vec{b}$ over a simulated time of 0.1 ns while allowing the white region to relax according to Newton's equation of motion and a constant temperature of 10 K. As mentioned above, for a partial dislocation loop, the MD simulation is completed directly; and for a perfect dislocation loop, the MD simulation is completed in two stages to follow the low energy path of two partial dislocations. This MD simulation procedure creates a dislocation loop at a shear strain of approximately γ = b/t, where b is magnitude of the dislocation Burgers vector, and t is thickness of the sample. The system is then re-divided into three regions as shown in Fig. 6(b): the black (top and bottom) regions, and the white (remaining) region. Evolution of dislocation loops is examined in subsequent MD simulations at a high temperature of 900 K where the top and bottom surface black layers are further moved in positive and negative Burgers vector directions to increase the shear strain.

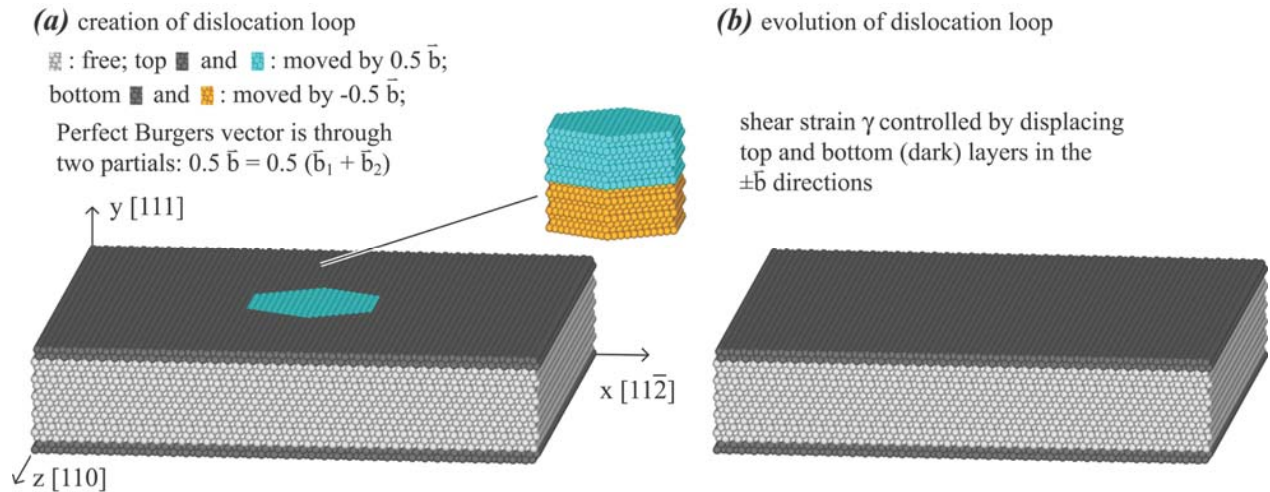

Fig. 6. Dislocation loop model. (a) MD simulation to create dislocation loop; and (b) MD simulation to evolve dislocation loop.

## A. Partial Dislocation Loop

A partial dislocation Burgers vector of $[11\bar{2}]a/6$ is used to create an initial dislocation loop using the procedure shown in Fig. 6(a). This procedure produces a shear strain of $\gamma_{xy} = 0.039$. Further shear strain is applied by moving the top and bottom black regions in the x and –x directions by 1.394 Å respectively using the procedure shown in Fig. 6(b). This corresponds to a total shear strain of $\gamma_{xy} = 0.078$. The equilibrium dislocation loop configurations obtained at 900 K temperature are examined in Fig. 7 by showing the top view of four consecutive planes above and blow the slip plane, where blue and gray circles indicated Cd and Te atoms respectively, and the yellow hexagon indicates the initial partial dislocation loop at $\gamma_{xy} = 0.039$. Figs. 7(a) and 7(b) are obtained using exactly the same conditions except that Cd and Te atoms are switched between the two figures, so that in Fig. 7(a), the right side of the loop is an α dislocation and the left side of the loop is a β dislocation, whereas in Fig. 7(b), the left side is α and the right side is β.

Because the movement of a partial dislocation leaves behind a stacking fault, the relaxed partial dislocation loop can be identified by the boundary of the two stacking patterns demonstrated in the figure. It can be seen from Fig. 7(a) that the hexagonal dislocation loop becomes asymmetric under the shear and in particular, the α dislocation moves by a far longer distance than the β dislocation. This is further verified in Fig. 7(b) where the asymmetry of the dislocation is flipped as a consequence of the flip of α and β dislocations. This simulation strongly indicates that α dislocations are much more mobile than β dislocations (despite their similar line energies in Table I), in good agreement with experiments conducted for many

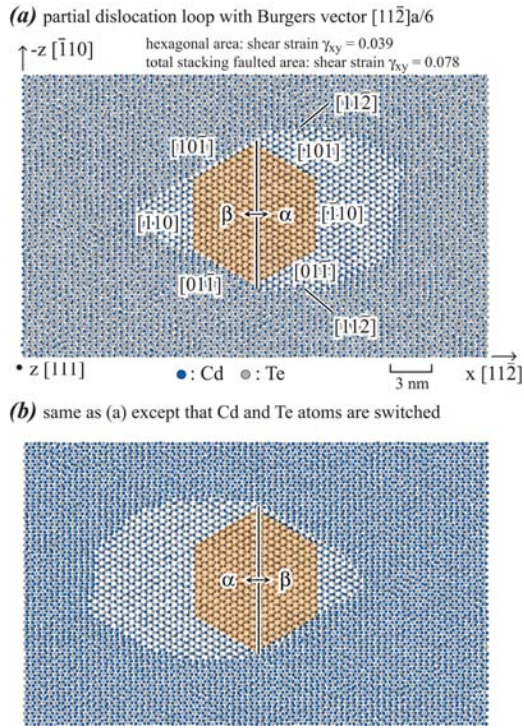

Fig. 7. Evolution of a $[11\bar{2}]a/6$ partial dislocation loop. (a) configurations at strains $\gamma_{xy} = 0.039$ and $\gamma_{xy} = 0.078$; and (b) the same as (a) except that Cd and Te atoms are switched.

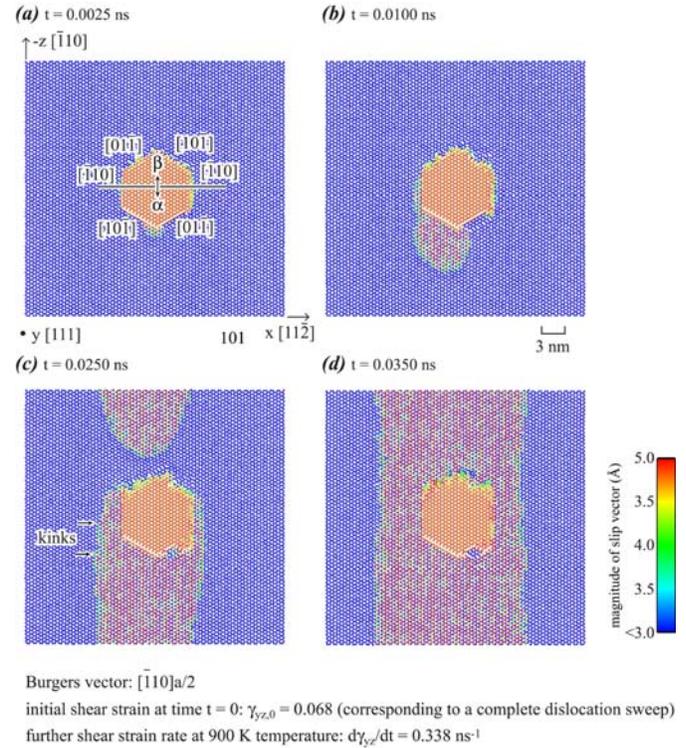

Fig. 8. Evolution of a $[\bar{1}10]/2$ perfect dislocation loop. (a) t = 0.0025 ns ; (b) t = 0.0100 ns; (c) t = 0.0250 ns; and (d) t = 0.0350 ns.

semiconductor compounds [22,23].

One interesting observation in Fig. 7 is that to accommodate the motion of the α segment ($\vec{\xi} = [\bar{1}10]$) of the dislocation loop, two segments ($\vec{\xi} = [11\bar{2}]$) are created. These segments are perpendicular to the α segment and are screw partials. This means that the motion of the mobile α partial dislocations can cause elongated screw partials. As a result, CdTe crystals may have a high density of such type of screw partials, in agreement with its low energy shown in Table I. On the other hand, the β type of edge partial dislocations disappear upon the strain. This corresponds well to the high energy of the edge partial in Table I.

*B. Perfect Dislocation Loop*

A perfect dislocation loop with a Burgers vector of $[\bar{1}10]a/2$ is created using the procedure described in Fig. 6(a). This procedure produces an initial shear strain of $\gamma_{yz,0} = 0.068$. By moving the top and bottom black regions in the z and –z directions at a constant speed of 1.2 nm/ns during an MD simulation at 900 K, as described in Fig. 6(b), the system is further deformed at a strain rate of $d\gamma_{yz}/dt = 0.338$ /ns. The configuration of the dislocation loop during the further strain is shown in Figs. 8(a) – 8(d) as a function of time, where the color scheme represents the magnitude of the slip vector developed previously to show dislocations [24]. Fig. 8(a) shows that shortly after further strain at t = 0.0025 ns ($\gamma_{yz} = 0.069$), the dislocation loop remains hexagonal but some portion of the α dislocation (the lower part of the [10$\bar{1}$] section) bows out. Fig. 8(b) shows that at t = 0.0100 ns ($\gamma_{yz}$ = 0.071), the dislocation bows out more. In Fig. 8(c) at 0.0250 ns ($\gamma_{yz}$ = 0.076), the α dislocation is seen to have moved a significant distance that causes it to pass through the low boundary and enter into the image from the top boundary under the periodic boundary condition. In contrast, little movement is seen for the β dislocation. Clearly, α perfect dislocations are also much more mobile than β perfect dislocations (despite their similar line energies shown in Table I), in good agreement with experiments [22]. Interestingly, Fig. 8(c) indicates that while screw dislocations also have higher mobility than β edge dislocations, they are much less mobile than α edge dislocations.

When the α dislocation meets the β dislocation at the top of the hexagon, they are annihilated, and the dislocation loop becomes two vertical screw dislocations running across the periodic dimension in the z direction, as can be seen in Fig. 8(d) at t = 0.035 ns ($\gamma_{yz}$ = 0.080). Further strain will cause the two dislocations to continuously move in the ± x directions until they completely sweep the entire area and are annihilated. Interestingly, we find that the dislocation motion is thermally activated and does not occur when the simulation is performed at a low temperature, say, 10 K. In addition, we find that the dislocations do not move as straight lines, but rather through the formation and rapid motion of kinks along the z direction as shown in Fig. 8(c). This is consistent with the traditional kink model of thermally activated motion of dislocations [25].

The observation that screw dislocations are much less mobile than α edge dislocations indicates that when the α sector of a dislocation loop moves, the screw dislocation sectors ($\bar{\bar{\xi}}$ = [$\bar{1}$ 10]) will elongate. As a result, screw dislocations can have a high density in the crystal. This is consistent with the low energies of screw dislocations shown in Table I. In addition, no dislocation dissociation is observed in Fig. 8, indicating that the perfect dislocations do not necessarily dissociate under non-equilibrium conditions. This is consistent with the discussions presented above.

## VII. DISLOCATION MOBILITIES

Dislocation mobility is important to understand because it directly controls dislocation network structures. For thermally-activated motion of dislocations, the dislocation velocity can be expressed as [26]

$$v = v_0 \cdot \exp\left(-\frac{Q - \tau \cdot \Omega}{kT}\right), \tag{1}$$

where Q and Ω are respectively activation energy and activation volume of dislocation motion, τ is shear stress, k is Boltzmann constant, T is temperature, and $v_0$ is a constant representing saturation dislocation velocity for barrierless (very high temperature or shear stress) dislocation motion. Here Q and Ω are the primary parameters quantifying dislocation mobility.

The discussion presented above already indicates several distinctive dislocation types that may determine dislocation

414structures. First, α and β partial and perfect dislocations have a strong edge component and they can contribute to the climb motion responsible for the formation of small angle grain boundaries. α dislocations are the most mobile and this may result in special structural features of dislocation networks. Screw partial and perfect dislocations can also be important. These screw dislocations are expected to have a high density as they have low energies and can be created by the motion of α dislocations. The objective of the present work is not to tabulate exhaustively the mobilities of a variety of dislocations. Hence, we focus to calculate activation energy barrier Q and activation volume Ω for α and β edge type of perfect dislocations. Such studies will allow us to fully explore the method for calculating Q and Ω and to validate the results using the known experimental relative mobilities of α and β dislocations. In addition, edge dislocations play an important role in dislocation network formation due to their climb motion. The extension of the method to other dislocations that can provide complete inputs for simulations of dislocation network structure evolution using larger scale (e.g., dislocation dynamics [27]) models will be performed in future work.

The activation energy barriers of dislocation motion are often directly calculated from energy profiles using the nudged elastic band method [28-31]. However, stop-action observation of the dislocation migration case shown in Fig. 8 indicates that the thermally activated motion of dislocations proceeds through a variety of different paths. Without requiring an explicit treatment of migration paths, we perform MD simulations of dislocation motion at a variety of temperatures and shear stresses to fit an apparent activation energy barrier and activation volume directly from (1). The activation energy determined this way pertains to a free energy barrier that incorporates the multiple path (entropy) effects. In addition, this approach provides a direct verification of (1), and hence, the thermally activated dislocation mechanism.

The model for performing dislocation motion simulations is shown in Fig. 9. The initial crystal contains 144 ($\bar{1}10$) planes in the x direction, 36 (111) planes in the y direction, and 24 ($11\bar{2}$) planes in the z direction. The system employs periodic boundary conditions in x and z directions and a free boundary condition in y direction. First, the initial crystal is divided into different regions as shown in Fig. 9(a): the black (left and right) regions, the blue (middle upper) region, the orange (middle lower) region, and the white (remaining) region. Since we only explore glide dislocations, we assume that the boundary between the blue and orange regions is between A and B planes as defined in Figs. 1(a) and 2(a). To create dislocations with a perfect Burgers vector of $\vec{b} = \vec{b}_1 + \vec{b}_2$, an MD simulation is performed where the black regions are fixed, the white regions are relaxed, and the blue and orange regions are first uniformly moved by $0.5\vec{b}_1$ and $-0.5\vec{b}_1$ respectively over a simulated time of 0.001 ns, and then moved by $0.5\vec{b}_2$ and $-0.5\vec{b}_2$ respectively over another simulated time of 0.001 ns. The crystal is then cut in half in the x direction as shown in Fig. 9(b) with a corresponding adjustment of the periodic length so that the two halves remain periodic in the x direction. This procedure essentially creates two crystals, each containing one dislocation. In particular, we use $\vec{b} =$ $[\bar{1}10]a/2$, $\vec{b}_1 = [\bar{1}2\bar{1}]a/6$, and $\vec{b}_2 = [\bar{2}11]a/6$ to create two crystals, one containing an α dislocation, and the other one

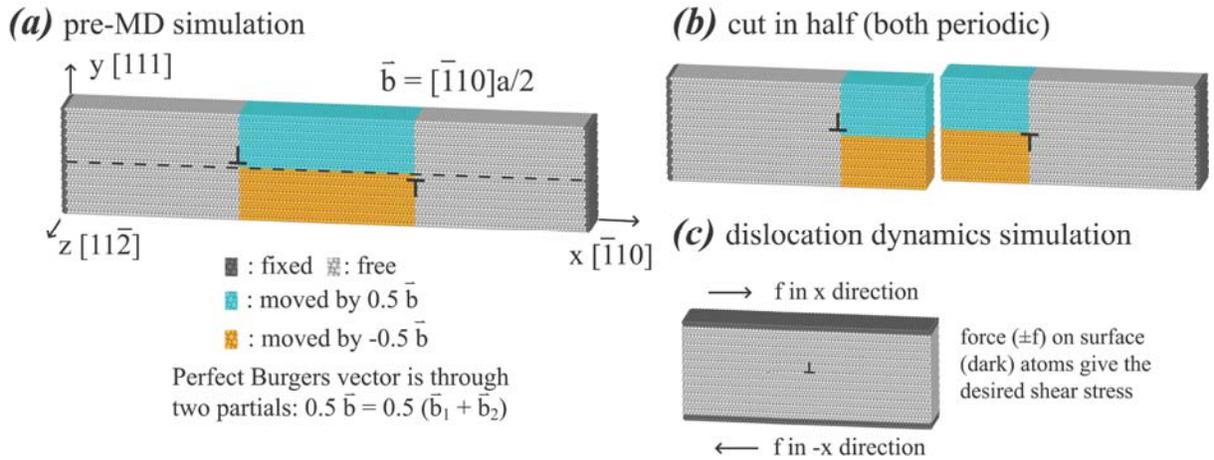

Fig. 9. Edge dislocation mobility model. (a) MD simulation to create α and β dislocations; (b) isolate out α and β dislocations; and (c) MD simulation to evolve dislocation under shear stress $\tau_{xy}$.

containing a β dislocation.

For dislocation motion simulations, the (half) system is divided into three regions: the black (top and bottom) regions and the white (remaining) region. By adding corresponding forces to the atoms in the top and bottom regions in +x and –x directions respectively, a desired shear stress $\tau_{xy}$ can be applied. Molecular dynamics simulations under npt (constant atom number, pressure, and temperature) condition are then performed to study migration of dislocations at a variety of stresses and temperatures. We find that in all of these simulations, dislocation splitting is very small, further verifying the discussions presented above that perfect dislocations do not necessarily dissociate. Assuming that the slip vector [24] reaches a maximum at the dislocation core, dislocation position is calculated as a function of time. Results obtained for α and β dislocations at selected stresses and temperatures are shown in Fig. 10. It can be seen from Fig. 10 that all the dislocation vs. time curves approximately fall on straight lines. This means that the data can provide reliable information about the steady-state dislocation velocity. The slope of these straight lines changes with stress and temperature, indicating that the simulated conditions capture the thermally

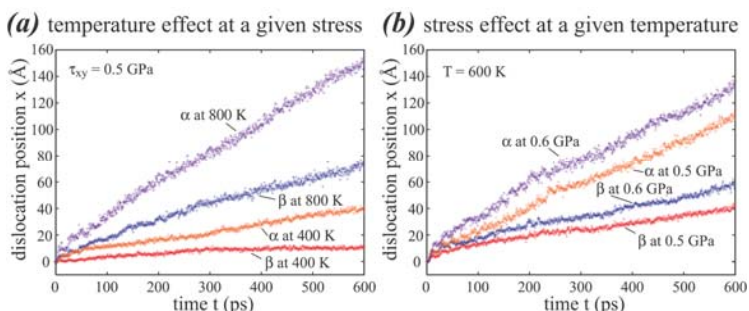

Fig. 10. Dislocation position as a function of time. (a) effect of temperature at a given stress; and (b) effect of stress at a given temperature.

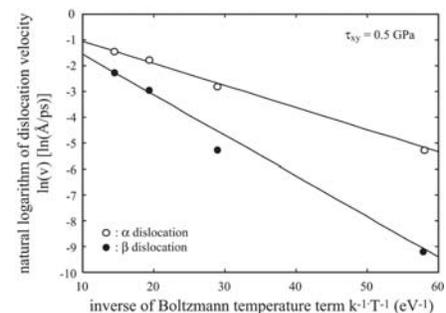

Fig. 11. Arrhenius relation between dislocation velocity and temperature.

activated mechanisms of dislocation motion.

Fig. 10(a) shows the effect of temperature on dislocation motion at a constant shear stress of 0.5 GPa. Increasing the temperature is seen to increase the dynamics, consistent with the thermal activation mechanism. At the same temperature, α dislocation is seen to move faster than β dislocation, further verifying the results seen in Figs. 7 and 8. Fig. 10(b) shows the effect of shear stress at a constant temperature of 600 K. Here increasing the stress also correctly increases the dislocation velocity, and the α dislocation is again validated to move faster than the β dislocation.

The steady state velocity can be easily calculated from the slope of the curves shown in Fig. 10. The results obtained at a constant shear stress of $\tau_{xy}$ = 0.5 GPa are plotted as ln(v) vs. $(kT)^{-1}$ curves in Fig. 11. Here the unfilled and filled circles represent MD data for α and β dislocations respectively, and the lines are fitted to the data using (1). Fig. 11 indicates that the MD data falls fairly well on straight $\ln(V)$ vs. $(kT)^{-1}$ lines, validating that the method indeed capture well the thermally activated dislocation migration. Through the fitting, we find that the activation energy and activation volume are Q = 0.14 eV and Ω = 17 $Å^3$ for the α dislocation and Q = 0.27 eV and Ω = 38 $Å^3$ for the β dislocation. This again verifies that in the thermally activated regime (e.g., $\tau_{xy}$ << 1 GPa), α dislocation is more mobile than β dislocation.

The thermally activated plastic deformation of CdTe has been experimentally analyzed below room temperature [32]. Activation energy and activation volume are assumed to decrease with increasing stress in the experiments. Under this assumption, the fitted experimental activation energy and activation volume are approximately 0.8 eV and 5500 $Å^3$ respectively at a small stress of 0.005 GPa, and fall to < 0.5 eV and < 2250 $Å^3$ respectively at a stress of 0.010 GPa. Here we assume that Q and Ω are constant and fitted those using MD data at high stresses (e.g., 0.5 GPa). In addition, simulations target only the Peierls potential barrier mechanism of specific (α and β perfect edge) dislocations, whereas, experiments may include point defect obstacle mechanisms and the contributions from other dislocations [32]. As a result, it is expected that the simulated activation energy and activation volume are smaller than their experimental counterparts.

## VIII. CONCLUSION

Our BOP-based MD simulations of dislocations in CdTe crystals lead to the following conclusions:

1) α dislocations move much faster than β dislocations, in good agreement with the well-known experiments for semiconductor compounds. This, along with previous demonstration that our BOP captures properties of many phases and predicts crystalline growth during MD simulations of chemical vapor deposition [17,19] and melt-growth [18], provides strong validation that the BOP correctly captures the physics of dislocation motion in CdTe. The BOP-based MD simulations, therefore, can provide high-fidelity data to construct large scale (e.g., dislocation dynamics [27]) models that can simulate the evolution of dislocation cell structures.

2) For partial, perfect, and shuffle sets, screw dislocations always have the lowest energies. Dissociated glide dislocations always have lower energies than perfect glide dislocations. However, perfect glide dislocations do not necessarily dissociate due to energy barrier of dissociation. In particular, the splitting distance of perfect glide dislocations is seen to remain very small in our dynamic MD simulations.

3) The acute angle of the second lowest energy (shuffle, glide partial, glide perfect) dislocations all equals $30^o$ (e.g., $\theta = 30^o$, $150^o$, $330^o$).

4) Screw dislocations are much less mobile than $\alpha$ edge dislocations. As a result, the screw dislocation sectors perpendicular to the $\alpha$ sector are elongated when an $\alpha$ sector of a dislocation loop moves. This suggests that screw dislocations are likely to have a higher density than edge dislocations, in good agreement with the low screw dislocation energies.

5) Without calculating energy profile explicitly, we show that MD simulations of dislocation motion at a variety of stresses and temperatures can be used to effectively deduce activation free energy barrier and activation volume of thermally activated glide of dislocations. In particular, we found activation free energy barriers of 0.14 eV and 0.27 eV and activation volumes of 17 $Å^3$ and 36 $Å^3$, respectively for $\alpha$ and $\beta$ edge type of glide dislocations.

## IX. Acknowledgment

This work is supported by the DOE/NNSA Office of Nonproliferation Research and Development, Proliferation Detection Program, Advanced Materials Portfolio. Sandia National Laboratories is a multi-program laboratory managed and operated by Sandia Corporation, a wholly owned subsidiary of Lockheed Martin Corporation, for the U.S. Department of Energy's National Nuclear Security Administration under contract DE-AC04-94AL85000.